\documentclass[twocolumn,showpacs,preprintnumbers,amsmath,amssymb]{revtex4}

\newcommand{\be}{\begin{equation}}
\newcommand{\ee}{\end{equation}}

\newcommand{\vspacecases}{\vspace{0.6 pc}}

\def\A{{\cal A}}
\def\B{{\cal B}}
\def\C{{\cal C}}
\def\F{{\cal F}}
\def\G{{\cal G}}
\def\H{{\cal H}}
\def\I{{\cal I}}
\def\M{{\cal M}}
\def\S{{\cal S}}
\def\V{{\cal V}}

\def\IM{{\int\limits_{M}}}
\def\b{\flat}
\def\#{\sharp}
\def\pd{\partial}

\def\Eq#1{(\ref{#1})}

\def\g{{\textstyle {\rm{g}}}}
\def\p{{\textstyle {\rm{p}}}}
\def\k{{\textstyle {\rm{k}}}}
\def\h{{\textstyle {\rm{h}}}}

\def\dmug{{\textstyle {\rm{d}}}\mu(\g)}
\def\tr{{\textstyle {\rm{tr}}}}
\def\div{{\textstyle {\rm{div}}}}

\def\half{{\textstyle {1\over2}}}

\begin{document}


\title{Is nonrelativistic gravity possible?}
\author{A.A.~Kocharyan}
\email{armen.kocharyan@sci.monash.edu.au}
\affiliation{%
School of Mathematical Sciences, Monash University, Clayton 3800, Australia
}%
\date{\today}

\begin{abstract} We study nonrelativistic gravity using the Hamiltonian formalism. For the dynamics of general relativity (relativistic gravity) the formalism is well known and called the Arnowitt-Deser-Misner (ADM) formalism. We show that if the lapse function is constrained correctly, then nonrelativistic gravity is described by a  consistent Hamiltonian system. Surprisingly, nonrelativistic gravity can have solutions identical to relativistic gravity ones. In particular, (anti-)de Sitter black holes of Einstein gravity and IR limit of Ho\v{r}ava gravity are locally identical.
\end{abstract}


\maketitle

\section{Introduction}

We use the Hamiltonian formalism \cite{Ref:ADM}, \cite{Ref:FiMa79a}, \cite{Ref:FiMa79b}, \cite{Ref:AAK} for the dynamics of nonrelativistic gravity in Wheeler-DeWitt superspace \cite{Ref:DeWitt}. The formalism leads naturally to the study of consistency of the nonrelativistic gravity. The equations of the rate of change of energy and momentum are computed. As is well known, the relativistic theory is characterised by identically zero energy rather than just the total integrated energy being zero \cite{Ref:Misner}, \cite{Ref:FiMa72}. A question arises: Can one generalise nonrelativistic theories and recover an identically zero energy condition? In other words: Can one generalise the lapse function from being a function of time only to a function of space and time? We show that the answer is negative, unless a very strong consistency condition is satisfied. Thus, generically, the lapse function of consistent nonrelativistic theories must be time dependent only. 

The approach is applicable to Ho\v{r}ava's recently proposed theory of gravity \cite{Ref:H1}, \cite{Ref:H2}. In particular, we show that there are no new (anti-)de Sitter black hole solutions. In fact, the theory has the same solutions as Einstein gravity in empty and flat space if $\lambda=1$.

\section{Nonrelativistic gravity}
\subsection{Superspace}
Let $M$ be an oriented without boundary smooth d-dimensional manifold. 
Let $S_2(M)$ denote the space of all smooth symmetric two tensors on $M$ and 
let $\M\subset S_2(M)$ be the manifold of positive definite Riemannian metrics on $M$. The tangent bundle of $\M$ is
$$
T\M=\M\times S_2(M).
$$
Let $S^2_d(M)$ be the space of all symmetric two contravariant tensor densities on $M$. The cotangent bundle of $\M$ is
$$
T^*\M=\M\times S^2_d(M).
$$
We have a natural pairing between $T\M$ and $T^*\M$ given by
$$
\langle \pi,\k \rangle = \IM\pi\cdot\k
=\IM\pi^{ab}\k_{ab}=\IM\dmug\p^{ab}\k_{ab},
$$
where $\pi\in T^*\M$, $\k\in T\M$, $\pi=\mbox{p}\dmug$, 
$\dmug=(\det\mbox{g})^{1/2} d\mbox{x}^1\wedge\cdots\wedge d\mbox{x}^{\rm{d}}$,
and $a,b=1,\ldots,\rm{d}$.

The DeWitt metric on $\M$ is given by \cite{Ref:DeWitt}
$$
G(\k,\k)=\IM\G(\k,\k)=\IM\dmug[\k\cdot\k-\lambda\ \tr(\k)\tr(\k)],
$$
where 
$\lambda$ is a constant, $\tr(\k)=\g^{ab}\k_{ab}$, 
$(\k\times\k)_{ab}=\k_{ac}\g^{cd}\k_{db}$, and 
$\k\cdot\k=\tr(\k\times\k)$.
The metric $G$ has an inverse metric $G^{-1}$ given by
$$
G^{-1}(\pi,\pi)=\IM\G^{-1}(\pi,\pi)
=\IM\dmug[\p\cdot\p-\tilde\lambda\ \tr(\p)\tr(\p)],
$$
where 
$$
\tilde\lambda=\frac{\lambda}{\lambda\rm{d}-1},\quad
\lambda\ne\frac{1}{\rm{d}}.
$$

\subsection{Hamiltonian formalism}
We investigate a dynamical system on $T\M$ given by an invariant action
\be
S=\int dt\IM N\left[\G(\k,\k)-\V(\g)\right],
\ee
where
$$
\k_{ab}
=\frac{1}{2N}\left(\frac{\pd}{\pd t}\g_{ab}-X_{a|b}-X_{b|a}\right)
=\frac{1}{2N}\left[\dot{\g}_{ab}-(L_X\g)_{ab}\right],
$$
$X$ (shift vector field) is a time dependant vector field on $M$, 
$N$ (lapse function) is a function of $t$ only, i.e. $N(t)$ is a constant function in the space of real-valued functions $\mathfrak{F}(M)$, 
$L_X$ is the Lie derivative, and the potential $\V(\g)\in\mathfrak{F}_d(M)$ is a scalar density.

The canonical momenta conjugate to $\g_{ab}$ are
$$
\pi^{ab}=\p^{ab}\dmug=\frac{\delta S}{\delta\dot\g_{ab}}
=(\k^{ab}-\lambda\ \tr(\k)\g^{ab})\dmug,
$$
and the Hamiltonian is 
\be\label{Eq:Ham}
H(\g,\pi)=\IM N \H(\g,\pi)+X\cdot\I(\g,\pi),
\ee
where 
\begin{align*}
\H(\g,\pi)&=\G^{-1}(\pi,\pi)+\V(\g),\\
\I(\g,\pi)&=2\delta\pi=-2\pi^b{}_{a|b},\\
X\cdot\I(\g,\pi)&=X^a\I_a(\g,\pi).
\end{align*}

Hamiltonian equations have the following form \cite{Ref:FiMa79a}, \cite{Ref:FiMa79b}:
\be\label{Eq:HamGen}
\begin{cases}
{\displaystyle\frac{\pd\g}{\pd t}}=2N\G_\b(\pi)+L_X\g,\vspacecases\\
{\displaystyle\frac{\pd\pi}{\pd t}}=N\S_\g(\pi,\pi) +\F(\g)\cdot N +L_X\pi,
\end{cases}
\ee

where
\begin{align*}
\G_\b(\pi)\cdot\pi&=\G^{-1}(\pi,\pi),\\
\S_\g(\pi,\pi)&=-2[\p\times\p -\tilde\lambda(\tr\p)\p]\dmug
+\half\g^{-1}\G^{-1}(\pi,\pi),\\
\F(\g)\cdot N&=-N\pd_\g\V(\g,\Gamma) -\B^*\cdot N.
\end{align*}
$\B$ and its adjoint map $\B^*$
\begin{align*}
&\B:T\M\to\mathfrak{F}_d(M):\h\mapsto\B\cdot\h,\\
&\B^*:\mathfrak{F}(M)\to T^*\M:N\mapsto\B^*\cdot N
\end{align*}
are defined by
\begin{align*}
\B\cdot\h&=D_\Gamma\V(\g,\Gamma)\cdot(D_\g\Gamma(\g)\cdot\h),\\
\IM N (\B\cdot\h)&=\IM (\B^*\cdot N)\cdot\h.
\end{align*}
Here we follow \cite{Ref:FiMa79a} and consider the potential $\V$ as a function of the undifferentiated metric coefficients $\g$ that do not appear in the Christoffel symbols $\Gamma$, and of the Christoffel symbols, and we write $\V(\g,\Gamma)$.

\subsection{Constraints}
The invariance of the Hamiltonian with respect to the
spatial diffeomorphisms implies the following \cite{Ref:FiMa79a}:
$$
0=\IM\pi\cdot L_X\g = \IM X\cdot\I,
$$
for an arbitrary vector field $X$. Therefore, we have the following conservation law (constraint) 
\be\label{Eq:I}
\I=0.
\ee
Then from \Eq{Eq:Ham} we get
\be\label{Eq:IH}
\IM\H=0,
\ee
but not necessarily a much stronger constraint
\be\label{Eq:H}
\H=0,
\ee 
as in relativistic gravity. As is well known \cite{Ref:Misner}, \cite{Ref:FiMa72}, in any topologically invariant theory \Eq{Eq:H} holds rather than just \Eq{Eq:IH}.

But, is it possible to impose \Eq{Eq:H} on nonrelativistic gravity? In order to answer this question let us compute the rate of change of $\H$ and $\I$ along a solution of \Eq{Eq:HamGen} for general $N(\rm{x},t)$ and $X(\rm{x},t)$. It is straightforward to show that (cf. \cite{Ref:FiMa79a})
\begin{align}\label{Eq:dHdI}
&
\begin{cases}
{\displaystyle\frac{d\H}{dt}}=\A_N +L_X\H,\vspace{0.6 pc}\\
{\displaystyle\frac{d\I}{dt}}=(dN)\H+L_X\I,\\
\end{cases}
\intertext{where}
&
\A_N(\g,\pi)
=2\G^{-1}(\B N-\B^*\cdot N,\pi).
\end{align}
Incidentally, \Eq{Eq:dHdI} is equivalent to the Dirac canonical commutation relations (cf. \cite{Ref:Dirac}, \cite{Ref:FiMa79a}, \cite{Ref:FiMa79b}).

Let us define \cite{Ref:FiMa79b}
\begin{align*}
\C_\H&=\{(\g,\pi)\in T^*\M\ |\ \H(\g,\pi)=0\},\\
\C_\I&=\{(\g,\pi)\in T^*\M\ |\ \I(\g,\pi)=0\},\\
\C&=\C_\H\cap\C_\I\\
&=\{(\g,\pi)\in T^*\M\ |\ \H(\g,\pi)=0,\ \I(\g,\pi)=0\}.
\end{align*}
If $(\g(0),\pi(0))\in\C$, then we have $(\g(t),\pi(t))\in\C_\I$ for all $t$ for which the solution exists, but $(\g(t),\pi(t))\in\C$ for all $t$ if and only if the restriction of $\A_N$ to $\C\subset T^*\M$ vanishes, i.e. the following condition holds for all $N$
\be\label{Eq:A}
\A_N(\g(t),\pi(t))\big|_\C=0.
\ee
If one assumes that $N$ is a function of $\rm{x}$ and $t$ for a nonrelativistic theory, then the theory will be consistent if and only if \Eq{Eq:A} holds. This is a very strong condition. By definition we have 
$$
\IM N\A_N=0.
$$ 
However, \Eq{Eq:A} does not hold for all $N$ and a general potential $\V(\g)$. We know one theory (possibly the only one if $\lambda\ne 1/d$), that of general relativity satisfying the condition. If \Eq{Eq:A} does not hold, then the Hamiltonian system is not consistent. Hence, \Eq{Eq:H} cannot be imposed and one has to consider $N$ as a function of $t$ only. In that case \Eq{Eq:dHdI} can be written in the following form:
\begin{align}\label{Eq:dHdIN}
&
\begin{cases}
{\displaystyle\frac{d\H}{dt}}=N\A +L_X\H,\vspacecases\\
{\displaystyle\frac{d\I}{dt}}=L_X\I,\\
\end{cases}
\intertext{where}
&\A(\g,\pi)=2\G^{-1}(\B-\B^*\cdot 1,\pi).
\end{align}
Thus, it is obvious that nonrelativistic gravity is possible, provided one considers a time only dependant lapse function, a projectable function (see \cite{Ref:H2}). If one generalises the lapse function, then the only meaningful, consistent theory is Einstein gravity.

However, if \Eq{Eq:A} does not hold for all solutions it can hold for specific solutions. Indeed, there could exist solutions with $\A(\g(t),\pi(t))\big|_\C=0$, then $\H(\g(t),\pi(t))=0$ and $\I(\g(t),\pi(t))=0$. These types of solutions would mimic relativistic ones. They will be called Lorentz symmetry recovering (LSR) solutions.

\subsection{Examples}
Let us consider some important (non)relativistic theories.

{\bf Einstein gravity}.  For the relativistic potential
$$
\V(\g)=(-R+2\Lambda)\dmug,
$$
with arbitrary $\lambda$ we have
\begin{align}
\F^{ab}&=-\left(R^{ab}-\half R\g^{ab} +\Lambda\g^{ab}\right)\dmug,\nonumber\\
\intertext{and}
\label{Eq:AE}
\A_N(\g,\pi)&=N^{-1}\div(N^2\I) -2N\frac{\lambda-1}{\lambda\,\rm{d}-1}\Delta\tr\pi,
\end{align}
where $\div Y=Y^a{}_{|a}$, and $\Delta f=-\g^{ab}f_{|ab}$. Thus, we see that $\lambda=1$ and $\lambda=1/\rm{d}$ are critical values as noted in \cite{Ref:H1}, \cite{Ref:H2}. Theories with $\lambda\ne1$ are very different from Einstein gravity, because of the last term in \Eq{Eq:AE}. The DeWitt metric's dependence on $\lambda=1$ is crucial too. If $\lambda=1$, then $\A_N(\g,\pi)\big|_\C=0$ and full relativistic gravity is recovered. Therefore, one is free to choose a space and time dependent lapse function.

{\bf Ho\v{r}ava gravity} \cite{Ref:H1}, \cite{Ref:H2}. We consider a more general potential
\begin{align*}
\V(\g)&=\left(\alpha_0+\alpha_1R+\alpha_2R^2\right.\\
&+\alpha_3R_{ab}R^{ab}
+\alpha_4\ \epsilon^{abc}R_{ad}R^d{}_{b|c}\\
&\left.+\,\alpha_5
\left[R_{ab|c}R^{ab|c}-R_{ab|c}R^{ac|b}-\frac{1}{8}R_{|a}R^{|a}\right]\right)\dmug.
\end{align*}
For simplicity, we assume that $\lambda=1$ and the spatial metric is flat $R_{ab}=0$, and then it is trivial to show that all solutions are LSR ones. Moreover, there is a bijection between solutions of Ho\v{r}ava and Einstein gravity. In particular, for a spherically symmetric metric, all solutions are locally equivalent to the Schwarzschild-Kottler solution in Lema\^{i}tre coordinates \cite{Ref:Lem}. For example, for $m>0$ and $\Lambda>0$, we have
$$
^4\g=-dt^2+\left(\frac{2m}{r}+\frac{1}{3}\Lambda r^2\right)d\rho^2 +r^2(d\theta^2+\sin^2\theta d\phi^2),
$$
where
$$
r^3(\rho,t)
=\frac{6m}{\Lambda}\ \sinh^2\left(\frac{\sqrt{3\Lambda}}{2}(\rho-t)\right).
$$
Thus, there is no ``new'' (A)dS black hole solutions in Ho\v{r}ava gravity. One will find new solutions if one considers a space and time dependent lapse function, but then the theory becomes inconsistent. However, nonflat geometries are not necessarily LSR solutions.

\section{Conclusions}
The Hamiltonian formalism is used to study nonrelativistic gravity. The evolution \Eq{Eq:dHdI} for $\H$ and $\I$ is derived and a consistency condition \Eq{Eq:A} is proposed. It is shown that if one considers a time only dependant lapse function, then nonrelativistic gravity is possible and described by a consistent Hamiltonian system. A typical nonrelativistic gravity will be an inconsistent theory if we assume a space and time dependant lapse function. One could conjecture that only Einstein gravity is consistent with a space and time dependant lapse function if $\lambda=1$.  The other possibility is Ho\v{r}ava gravity if $\lambda=1/\rm{d}$ (see \cite{Ref:H1}, \cite{Ref:H2}). 

The results of the paper can be extended to include field theories coupled to gravity. One is tempted to extend the approach and investigate the nonrelativistic Wheeler-DeWitt equation \cite{Ref:DeWitt}
$$
\left[\IM\G^{-1}\left(\frac{\delta}{\delta\g},\frac{\delta}{\delta\g}\right) -\V(\g)\right]\Psi(^3\mathfrak{g})=0.
$$
All of these directions will be investigated in further study and hopefully a more important question, ``Is physically meaningful nonrelativistic gravity possible?'' will be answered.

Similar issues with different assumptions are discussed in \cite{Ref:ChNPS}, \cite{Ref:LP}, \cite{Ref:SVW}.

{\it Note added.}--While this work was being prepared for submission, we became aware of \cite{Ref:SM} where similar questions are addressed.


\begin{thebibliography}{00}

\bibitem{Ref:ADM}
R.L.~Arnowitt, S.~Deser, and C.W.~Misner, Phys.~Rev. {\bf 117}, 1595 (1960).

\bibitem{Ref:FiMa79a}
A.E.~Fischer and J.E.~Marsden, in {\it Isolated Gravitating Systems in General Relativity}, edited by J.~Ehlers 
(North-Holland Publishing Co., Amsterdam, The Netherlands, 1979), p.~322.

\bibitem{Ref:FiMa79b}
A.E.~Fischer and J.E.~Marsden, in {\it General Relativity. An Einstein Centenary
Survey}, edited by S.W. Hawking and W. Israel
(Cambridge University Press, Cambridge, England, 1980) p.~138.

\bibitem{Ref:AAK}
A.A.~Kocharyan, Commun.~Math.~Phys. {\bf 143}, 27 (1991).

\bibitem{Ref:DeWitt}
B.S.~DeWitt, Phys.~Rev. {\bf 160}, 1113 (1967).

\bibitem{Ref:Misner}
C.W.~Misner, Rev.~Mod.~Phys. {\bf 29}, 497 (1957).

\bibitem{Ref:FiMa72}
A.E.~Fischer and J.E.~Marsden, J.~of~Math.~Phys. (N.Y.) {\bf 13}, 546 (1972).

\bibitem{Ref:H1}
P.~Ho\v{r}ava, J. High Energy Phys. {\bf 03}, 020 (2009).

\bibitem{Ref:H2}
P.~Ho\v{r}ava, Phys.~Rev.~D {\bf 79}, 084008 (2009).

\bibitem{Ref:Dirac}
P.~A.~M.~ Dirac, Phys.~Rev. {\bf 114}, 924 (1959).

\bibitem{Ref:Lem}
G.~Lema\^{i}tre, Ann.~Soc.~Sci.~Bruxelles, Ser.~1, {\bf A53}, 51 (1933).

\bibitem{Ref:ChNPS}
C.\,Charmousis,\,G.\,Niz,\,A.\,Padilla,\,and\,P.M.\,Saffin,\\ 
arXiv:hep-th:0905.2579.

\bibitem{Ref:LP}
M.~Li and Y.~Pang, arXiv:hep-th:0905.2751.

\bibitem{Ref:SVW}
T.P.~Sotiriou, M.~Visser, and S.~Weinfurtner,\\ arXiv:hep-th:0905.2798.

\bibitem{Ref:SM}
S.~Mukohyama, arXiv:hep-th:0905.3563.

\end{thebibliography}
\end{document}